# PATHWISE COORDINATE OPTIMIZATION

By Jerome Friedman,[1] Trevor Hastie,[2] Holger Höfling[3]
and Robert Tibshirani[4]

*Stanford University*

We consider "one-at-a-time" coordinate-wise descent algorithms
for a class of convex optimization problems. An algorithm of this kind
has been proposed for the $L_1$-penalized regression (lasso) in the liter-
ature, but it seems to have been largely ignored. Indeed, it seems that
coordinate-wise algorithms are not often used in convex optimization.
We show that this algorithm is very competitive with the well-known
LARS (or homotopy) procedure in large lasso problems, and that it
can be applied to related methods such as the garotte and elastic net.
It turns out that coordinate-wise descent does not work in the "fused
lasso," however, so we derive a generalized algorithm that yields the
solution in much less time that a standard convex optimizer. Finally,
we generalize the procedure to the two-dimensional fused lasso, and
demonstrate its performance on some image smoothing problems.

**1. Introduction.** In this paper we consider statistical models that lead
to convex optimization problems with inequality constraints. Typically, the
optimization for these problems is carried out using a standard quadratic
programming algorithm. The purpose of this paper is to explore "one-at-a-
time" coordinate-wise descent algorithms for these problems. The equivalent
of a coordinate descent algorithm has been proposed for the $L_1$-penalized
regression (lasso) in the literature, but it is not commonly used. Moreover,
coordinate-wise algorithms seem too simple, and they are not often used in
convex optimization, perhaps because they only work in specialized prob-
lems. We ourselves never appreciated the value of coordinate descent meth-
ods for convex statistical problems before working on this paper.

In this paper we show that coordinate descent is very competitive with
the well-known LARS (or homotopy) procedure in large lasso problems, can

Received May 2007; revised August 2007.

[1]Supported in part by NSF Grant DMS-97-64431.

[2]Supported in part by NSF Grant DMS-05-50676 and NIH Grant 2R01 CA 72028-07.

[3]Supported by an Albion Walter Hewlett Stanford Graduate Fellowship.

[4]Supported in part by NSF Grant DMS-99-71405 and NIH Contract N01-HV-28183.

*Key words and phrases.* Coordinate descent, lasso, convex optimization.

deliver a path of solutions efficiently, and can be applied to many other convex statistical problems such as the garotte and elastic net. We then go on to explore a nonseparable problem in which coordinate-wise descent does not work—the "fused lasso." We derive a generalized algorithm that yields the solution in much less time that a standard convex optimizer. Finally, we generalize the procedure to the two-dimensional fused lasso, and demonstrate its performance on some image smoothing problems.

A key point here: coordinate descent works so well in the class of problems that we consider because each coordinate minimization can be done quickly, and the relevant equations can be updated as we cycle through the variables. Furthermore, often the minimizers for many of the parameters don't change as we cycle through the variables, and hence, the iterations are very fast.

Consider, for example, the lasso for regularized regression [Tibshirani (1996)]. We have predictors $x_{ij}, j = 1, 2, \ldots, p$, and outcome values $y_i$ for the $i$th observation, for $i = 1, 2, \ldots, n$. Assume that the $x_{ij}$ are standardized so that $\sum_i x_{ij}/n = 0, \sum_i x_{ij}^2 = 1$. The lasso solves

$$\min_\beta \tfrac{1}{2} \sum_{i=1}^n \left( y_i - \sum_{j=1}^p x_{ij}\beta_j \right)^2$$

(1)

$$\text{subject to} \quad \sum_{j=1}^p |\beta_j| \le s.$$

The bound $s$ is a user-specified parameter, often chosen by a model selection procedure such as cross-validation. Equivalently, the solution to (1) also minimizes the "Lagrange" version of the problem

$$(2) \qquad f(\beta) = \tfrac{1}{2} \sum_{i=1}^n \left( y_i - \sum_{j=1}^p x_{ij}\beta_j \right)^2 + \gamma \sum_{j=1}^p |\beta_j|,$$

where $\gamma \ge 0$. There is a one-to-one correspondence between $\gamma$ and the bound $s$—if $\hat{\beta}(\gamma)$ minimizes (2), then it also solves (1) with $s = \sum_{j=1}^p |\hat{\beta}_j(\gamma)|$. In the signal processing literature, the lasso and $L_1$ penalization is known as "basis pursuit" [Chen et al. (1998)].

There are efficient algorithms for solving this problem for all values of $s$ or $\gamma$; see Efron et al. (2004), and the homotopy algorithm of [Osborne et al. (2000)]. There is another, simpler algorithm for solving this problem for a fixed value $\gamma$. It relies on solving a sequence of single-parameter problems, which are assumed to be simple to solve.

With a single predictor, the lasso solution is very simple, and is a soft-thresholded version [Donoho and Johnstone (1995)] of the least squares estimate $\hat{\beta}$:

$$(3) \qquad \hat{\beta}^{\text{lasso}}(\gamma) = S(\hat{\beta}, \gamma) \equiv \text{sign}(\hat{\beta})(|\hat{\beta}| - \gamma)_+$$

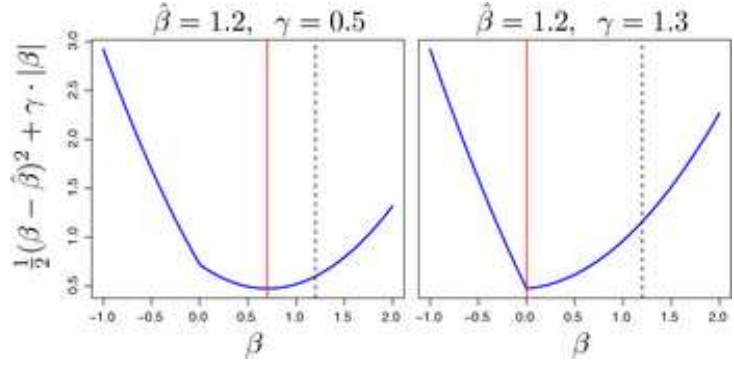

FIG. 1.  *The lasso problem with a single standardized predictor leads to soft thresholding. In each case the solid vertical line indicates the lasso estimate, and the broken line the least-squares estimate.*

$$(4) \qquad = \begin{cases} \hat{\beta} - \gamma, & \text{if } \hat{\beta} > 0 \text{ and } \gamma < |\hat{\beta}|, \\ \hat{\beta} + \gamma, & \text{if } \hat{\beta} < 0 \text{ and } \gamma < |\hat{\beta}|, \\ 0, & \text{if } \gamma \geq |\hat{\beta}|. \end{cases}$$

This simple expression arises because the convex-optimization problem (2) reduces to a few special cases when there is a single predictor. Minimizing the criterion (2) with a single standardized $x$ and $\beta$ simplifies to the equivalent problem

$$(5) \qquad \min_{\beta} \tfrac{1}{2}(\beta - \hat{\beta})^2 + \gamma|\beta|,$$

where $\hat{\beta} = \sum_i x_i y_i$ is the simple least-squares coefficient. If $\beta > 0$, we can differentiate (5) to get

$$(6) \qquad \frac{df}{d\beta} = \beta - \hat{\beta} + \gamma = 0.$$

This leads to the solution $\beta = \hat{\beta} - \gamma$ (left panel of Figure 1) as long as $\hat{\beta} > 0$ and $\gamma < \hat{\beta}$, otherwise 0 is the minimizing solution (right panel). Similarly, if $\hat{\beta} < 0$, if $\gamma < -\hat{\beta}$, then the solution is $\beta = \hat{\beta} + \gamma$, else 0.

With multiple predictors that are uncorrelated, it is easily seen that once again the lasso solutions are soft-thresholded versions of the individual least squares estimates. This is not the case for general (correlated) predictors. Consider instead a simple iterative algorithm that applies soft-thresholding with a "partial residual" as a response variable. We write (2) as

$$(7) \qquad f(\tilde{\beta}) = \tfrac{1}{2} \sum_{i=1}^n \left( y_i - \sum_{k \neq j} x_{ik}\tilde{\beta}_k - x_{ij}\beta_j \right)^2 + \gamma \sum_{k \neq j} |\tilde{\beta}_j| + \gamma|\beta_j|,$$

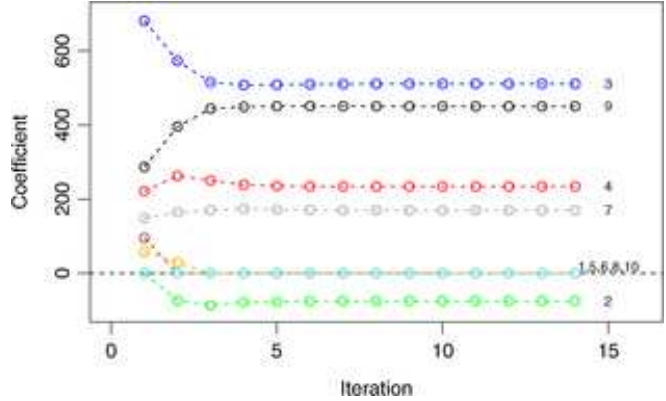

FIG. 2.  *Diabetes data: iterates for each coefficient from algorithm* (9). *The algorithm converges to the lasso estimates shown on the right side of the plot.*

where all the values of $\beta_k$ for $k \neq j$ are held fixed at values $\tilde{\beta}_k(\gamma)$. Minimizing w.r.t. $\beta_j$, we get

$$(8) \qquad \tilde{\beta}_j(\gamma) \leftarrow S\left(\sum_{i=1}^{n} x_{ij}(y_i - \tilde{y}_i^{(j)}), \gamma\right),$$

where $\tilde{y}_i^{(j)} = \sum_{k \neq j} x_{ik}\tilde{\beta}_k(\gamma)$. This is simply the univariate regression coefficient of the partial residual $y_i - \tilde{y}_i^{(j)}$ on the (unit $L_2$-norm) $j$th variable; hence, this has the same form as the univariate version (3) above. The update (7) is repeated for $j = 1, 2, \ldots, p, 1, 2, \ldots$ until convergence.

An equivalent form of this update is

$$(9) \quad \tilde{\beta}_j(\gamma) \leftarrow S\left(\tilde{\beta}_j(\gamma) + \sum_{i=1}^{n} x_{ij}(y_i - \tilde{y}_i), \gamma\right), \qquad j = 1, 2, \ldots p, 1, 2, \ldots.$$

Starting with any values for $\beta_j$, for example, the univariate regression coefficients, it can be shown that the $\tilde{\beta}_j(\gamma)$ values converge to $\hat{\beta}^{\text{lasso}}$.

Figure 2 shows an example, using the diabetes data from Efron et al. (2004). This data has 442 observations and 10 predictors. We applied algorithm (9) with $\gamma = 88$. It produces the iterates shown in the figure and converged after 14 steps to the lasso solution $\hat{\beta}^{\text{lasso}}(88)$.

This approach provides a simple but fast algorithm for solving the lasso, especially useful for large $p$. It was proposed in the "shooting" procedure of Fu (1998) and re-discovered by [Daubechies, Defrise and De Mol (2004)]. Application of the same idea to the elastic net procedure [Zhou and Hastie (2005)] was proposed by [Van der Kooij (2007)].

**2. Pathwise coordinatewise optimization algorithms.** The procedure described in Section 1 is a successive coordinate-wise descent algorithm for minimizing the function $f(\beta) = \frac{1}{2}\sum_i(y_i - \sum_j x_{ij}\beta_j)^2 + \lambda\sum_{j=1}^p|\beta_j|$. The idea is to apply a coordinate-wise descent procedure for each value of the regularization parameter, varying the regularization parameter along a path. Each solution is used as a warm start for the next problem.

This approach is attractive whenever the single-parameter problem is easy to solve. Some of these algorithms have already been proposed in the literature, and we give appropriate references. Here are some other examples:

- *The nonnegative garotte.* This method, a precursor to the lasso, was proposed by Breiman (1995) and solves

$$(10) \qquad \min_c \tfrac{1}{2}\sum_{i=1}^n\left(y_i - \sum_{j=1}^p x_{ij}c_j\hat{\beta}_j\right)^2 + \lambda\sum_{j=1}^p c_j \quad \text{subject to} \quad c_j \geq 0.$$

Here $\hat{\beta}_j$ are the usual least squares estimates (we assume that $p \leq n$). Using partial residuals as in (7), one can show that the the coordinate-wise update has the form

$$(11) \qquad c_j \leftarrow \left(\frac{\tilde{\beta}_j\hat{\beta}_j - \lambda}{\hat{\beta}_j^2}\right)_+,$$

where $\tilde{\beta}_j = \sum_{i=1}^n x_{ij}(y_i - \tilde{y}_i^{(j)})$, and $\tilde{y}_i^{(j)} = \sum_{k\neq j} x_{ik}c_k\hat{\beta}_k$.

- *Least absolute deviation regression and LAD-lasso.* Here the problem is

$$(12) \qquad \min_\beta \sum_{i=1}^n\left|y_i - \beta_0 - \sum_{j=1}^p x_{ij}\beta_j\right|.$$

We can write this as

$$(13) \qquad \sum_{i=1}^n|x_{ij}|\left|\frac{(y_i - \tilde{y}_i^{(j)})}{x_{ij}} - \beta_j\right|,$$

holding all but $\beta_j$ fixed. This quantity is minimized over $\beta_j$ by a weighted median of the values $(y_i - \tilde{y}_i^{(j)})/x_{ij}$. Hence, coordinate-wise minimization is just a repeated computation of medians. This approach is studied and refined by Li and Arce (2004).

The same approach may be used in the LAD-lasso [Wang et al. (2006)]. Here we add an $L_1$ penalty to the least absolute deviation loss:

$$(14) \qquad \min_\beta \sum_{i=1}^n\left|y_i - \beta_0 - \sum_{j=1}^p x_{ij}\beta_j\right| + \lambda\sum_{j=1}^p|\beta_j|.$$

This can be converted into an LAD problem (12) by augmenting the dataset with $p$ artificial observations. In detail, let **X** denote $n \times (p +$

1) model matrix (with the column of 1s for the intercept). The extra $p$ observations have response $y_i$ equal to zero, and predictor matrix equal to $\lambda \cdot (\mathbf{0} : \mathbf{I}_p)$. Then the above coordinate-wise algorithm for the LAD problem can be applied.

• *The elastic net.* This method [due to Zhou and Hastie (2005)] adds a second constraint $\sum_{j=1}^{p} \beta_j^2 \le s_2$ to the lasso (1). In Lagrange form, we solve

$$(15) \qquad \min_{\beta} \frac{1}{2} \sum_{i=1}^{n} \left( y_i - \sum_{j=1}^{n} x_{ij} \beta_j \right)^2 + \lambda_1 \sum_{j=1}^{p} |\beta_j| + \lambda_2 \sum_{j=1}^{p} \beta_j^2 / 2.$$

The coordinate-wise update has the form

$$(16) \qquad \tilde{\beta}_j \leftarrow \frac{S(\sum_{i=1}^{n} x_{ij}(y_i - \tilde{y}_i^{(j)}), \lambda_1)_+}{1 + \lambda_2}.$$

Thus, we compute the simple least squares coefficient on the partial residual, apply soft-thresholding to take care of the lasso penalty, and then apply a proportional shrinkage for the ridge penalty. This algorithm was suggested by Van der Kooij (2007).

• *Grouped lasso* [Yuan and Lin (2006)]. This is like the lasso, except variables occur in groups (such as dummy variables for multi-level factors). Suppose $X_j$ is an $N \times p_j$ orthonormal matrix that represents the $j$th group of $p_j$ variables, $j = 1, \ldots, m$, and $\beta_j$ the corresponding coefficient vector. The grouped lasso solves

$$(17) \qquad \min_{\beta} \left\| y - \sum_{j=1}^{m} X_j \beta_j \right\|_2^2 + \sum_{j=1}^{m} \lambda_j \|\beta_j\|_2,$$

where $\lambda_j = \lambda \sqrt{p_j}$. Other choices of $\lambda_j \ge 0$ are possible; this one penalizes large groups more heavily. Notice that the penalty is a weighted sum of $L_2$ norms (not squared); this has the effect of selecting the variables in groups. Yuan and Lin (2006) argue for a coordinate-descent algorithm for solving this problem, and show through the Karush–Kuhn–Tucker conditions that the coordinate updates are given by

$$(18) \qquad \tilde{\beta}_j \leftarrow (\|S_j\|_2 - \lambda_j)_+ \frac{S_j}{\|S_j\|_2},$$

where $S_j = X_j^T(y - \tilde{y}^{(j)})$, and here $\tilde{y}^{(j)} = \sum_{k \ne j} X_k \tilde{\beta}_k$.

• *The "Berhu" penalty.* This method is due to Owen (2006), and is another compromise between an $L_1$ and $L_2$ penalty. In Lagrange form, we solve

$$\min_{\beta} \frac{1}{2} \sum_{i=1}^{n} \left( y_i - \sum_{j=1}^{n} x_{ij} \beta_j \right)^2$$

(19)
$$+ \lambda \sum_{j=1}^{p} \Big[ |\beta_j| \cdot I(|\beta_j| < \delta) + \frac{(\beta_j^2 + \delta^2)}{2\delta} \cdot I(|\beta_j| \geq \delta) \Big].$$

This penalty is the reverse of a "Huber" function—initially absolute value, but then blending into quadratic beyond $\delta$ from zero. The coordinate-wise update has the form

(20)
$$\tilde{\beta}_j \leftarrow \begin{cases} S\Big( \sum_{i=1}^{n} x_{ij}(y_i - \tilde{y}^{(j)}), \lambda \Big), & \text{if } |\beta_j| < \delta, \\ \sum_{i=1}^{n} x_{ij}(y_i - \tilde{y}^{(j)})/(1 + \lambda/\delta), & \text{if } |\beta_j| \geq \delta. \end{cases}$$

This is a lasso-style soft-thresholding for values less than $\delta$, and ridge-style beyond $\delta$.

Tseng (1988) [see also Tseng (2001)] has established that coordinate descent works in problems like the above. He considers minimizing functions of the form

(21)
$$f(\beta_1, \ldots, \beta_p) = g(\beta_1, \ldots, \beta_p) + \sum_{i=j}^{p} h_j(\beta_j),$$

where $g(\cdot)$ is differentiable and convex, and the $h_j(\cdot)$ are convex. Here each $\beta_j$ can be a vector, but the different vectors cannot have any overlapping members. He shows that coordinate descent converges to the minimizer of $f$. The key to this result is the separability of the penalty function $\sum_{j=1}^{p} h_j(\beta_j)$, a sum of functions of each individual parameter. This result implies that the coordinate-wise algorithms for the lasso, the grouped lasso and elastic net, etc. converge to their optimal solutions.

Next we examine coordinate-wise descent in a more complicated problem, the "fused lasso" [Tibshirani, Saunders, Rosset, Zhu and Knight (2005)], which we represent in Lagrange form:

(22)
$$f(\beta) = \tfrac{1}{2} \sum_{i=1}^{n} \Big( y_i - \sum_{j=1}^{p} x_{ij}\beta_j \Big)^2 + \lambda_1 \sum_{j=1}^{p} |\beta_j| + \lambda_2 \sum_{j=2}^{p} |\beta_j - \beta_{j-1}|.$$

The first penalty encourages sparsity in the coefficients; the second penalty encourages sparsity in their differences; that is, flatness of the coefficient profiles $\beta_j$ as a function of the index set $j$. Note that $f(\beta)$ is strictly convex, and hence has a unique minimum.

The left panel of Figure 3 shows an example of an application of the fused lasso, in a special case where the feature matrix $\{x_{ij}\}$ is the identity matrix—this is called *fused lasso signal approximation*, discussed in the next section. The data represents Comparative Genomic Hybridization (CGH) measurements from two Glioblastoma Multiforme (GBM) tumors. The measurements are "copy numbers"—log-ratios of the number of copies of each gene in the tumor versus normal tissue. The data are displayed in the left panel, and the red line in the right panel represents the smoothed signal $\hat{\beta}$ from the fused lasso. The regions of the nonzero estimated signal can be used to call "gains" and "losses" of genes. Tibshirani and Wang (2007) report excellent results in the application of the fused lasso, finding that the method outperforms other popular techniques for this problem.

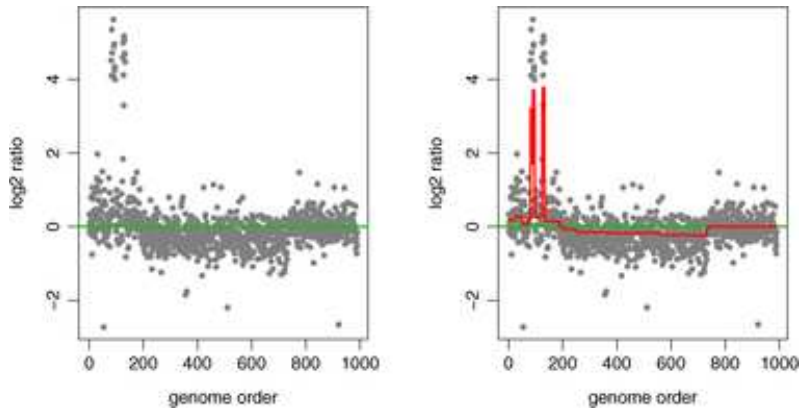

FIG. 3.   *Fused lasso applied to some Glioblastoma Multiforme data. The data are shown in the left panel, and the jagged line in the right panel represents the inferred copy number $\hat{\beta}$ from the fused lasso. The horizontal line is for $y = 0$.*

Somewhat surprisingly, coordinate-wise descent does not work for the fused lasso. Proposition 2.7.1 of Bertsekas (1999), for example, shows that every limit point of successive coordinate-wise minimization of a continuously differentiable function is a stationary point for the overall minimization, provided that the minimum is uniquely obtained along each coordinate. However, $f(\beta)$ is not continuously differentiable, which means that coordinate-wise descent can get stuck. Looking at Tseng's result, the penalty function for the fused lasso is not separable, and hence, Tseng's theorem cannot be applied in that case.

Figure 4 illustrates the difficulty. We created a fused lasso problem with 100 parameters, with the solutions for two of the parameters, $\beta_{63}$ and $\beta_{64}$, being equal to about $-1$. The top panels shows slices of the function $f(\beta)$ varying $\beta_{63}$ and $\beta_{64}$, with the other parameters set to the global minimizers. We see that the coordinate-wise descent algorithm has got stuck in a corner of the response surface, and is stationary under single-coordinate moves. In order to advance to the minimum, we have to move both $\beta_{63}$ and $\beta_{64}$ together.

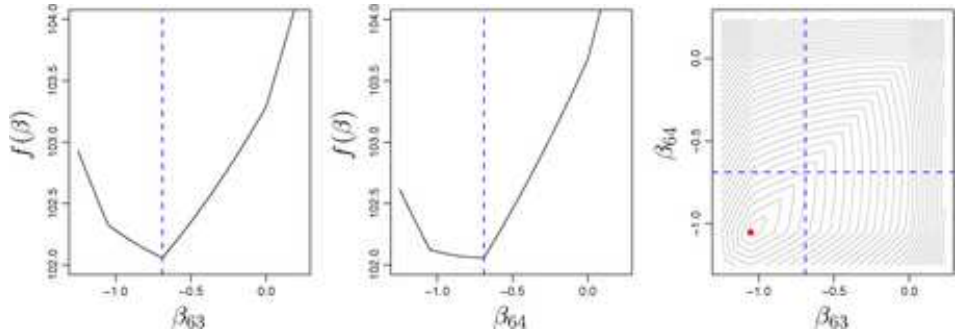

Fɪɢ. 4. *Failure of coordinate-wise descent in a fused lasso problem with 100 parameters. The optimal values for two of the parameters, $\beta_{63}$ and $\beta_{64}$, are both $-1.05$, as shown by the dot in the right panel. The left and middle panels shows slices of the objective function $f(\beta)$ as a function of $\beta_{63}$ and $\beta_{64}$, with the other parameters set to the global minimizers. The coordinate-wise minimizer over both $\beta_{63}$ and $\beta_{64}$ (separately) is $-0.69$, rather than $-1.05$. The right panel shows contours of the two-dimensional surface. The coordinate-descent algorithm is stuck at $(-0.69, -0.69)$. Despite being strictly convex, the surface has corners, in which the coordinate-wise procedure can get stuck. In order to travel to the minimum we have to move both $\beta_{63}$ and $\beta_{64}$ together.*

Despite this, it turns out that the coordinate-wise descent procedure can be modified to work for the fused lasso, yielding an algorithm that is much faster than a general quadratic-program solver for this problem.

**3. The fused lasso signal approximator.** Here we consider a variant of the fused lasso (22) for approximating one- and higher-dimensional signals, which we call the *fused-lasso signal approximator* (FLSA). For one-dimensional signals we solve

$$(23) \qquad \min_{\beta} f(\beta) = \tfrac{1}{2} \sum_{i=1}^{n} (y_i - \beta_i)^2 + \lambda_1 \sum_{i=1}^{n} |\beta_i| + \lambda_2 \sum_{i=2}^{n} |\beta_i - \beta_{i-1}|.$$

The measurements $y_i$ are made along a one-dimensional index $i$, and there is one parameter per observation. Later we consider images as well. In the special case of $\lambda_1 = 0$, the fused-lasso signal approximator is equivalent to a discrete version of the "total variation denoising" procedure [Rudin et al. (1992)] used in signal processing. We make this connection clear in Section 6. Thus, the algorithm that we present here also provides a fast implementation for total variation denoising.

Figure 5 illustrates an example of FLSA with 1000 simulated data points, and the fit is shown for $s_1 = 269.2, s_2 = 10.9$.

We now describe a modified coordinate-wise algorithm for the diagonal fused lasso (FLSA) using the Lagrange form (23). The algorithm can also be extended to the general fused lasso problem; details are given in the [Appendix](#). However, it is not guaranteed to give the exact solution for that problem, as we later make clear.

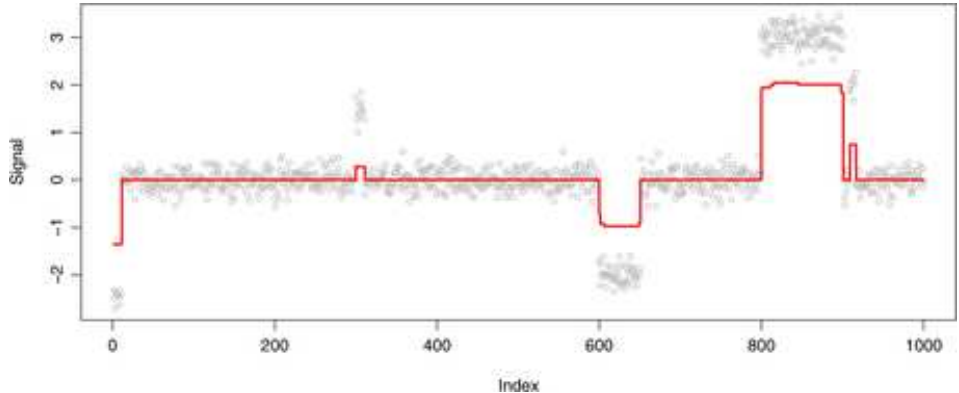

Fig. 5.   *Fused lasso solution in a constructed example.*

Our algorithm, for a fixed value $\lambda_1$, delivers a sequence of solutions corresponding to an increasing sequence of values for $\lambda_2$. First we make an observation that makes it easy to obtain the solution over a grid of $\lambda_1$ and $\lambda_2$ values:

PROPOSITION 1.   *The solutions to* (23) *for all* $(\lambda'_1 > \lambda_1, \lambda_2)$ *can be obtained by soft-thresholding the solution for* $(\lambda_1, \lambda_2)$.

This is proven in the [Appendix](#) for the FLSA, two-dimensional penalty fused lasso and even more general penalty functionals. Thus, our overall strategy for obtaining the solution over a grid is to solve the problem for $\lambda_1 = 0$ over a grid of values of $\lambda_2$, and then use this result to obtain the solution for all values of $\lambda_1$. However, for a single (especially large) value of $\lambda_1$, we demonstrate that it is faster to obtain the solution directly for that value of $\lambda_1$ (Table [2](#)). Hence, we present our algorithm for fixed but arbitrary values of $\lambda_1$.

Two keys for the algorithm are assumptions (A1) and (A2) below, stating that for fixed $\lambda_1$, small increments in the value of $\lambda_2$ can only cause pairs of parameters to fuse and they do not become unfused for the larger $\lambda_2$. This allows us to efficiently solve the problem for a path of $\lambda_2$ values, keeping $\lambda_1$ fixed.

The algorithm has three nested cycles:

*Descent cycle*: Here we run coordinate-wise descent for each parameter $\beta_j$, holding all the others fixed.
*Fusion cycle*: Here we consider the fusion of a neighboring pair of parameters, followed by coordinate-wise descent.

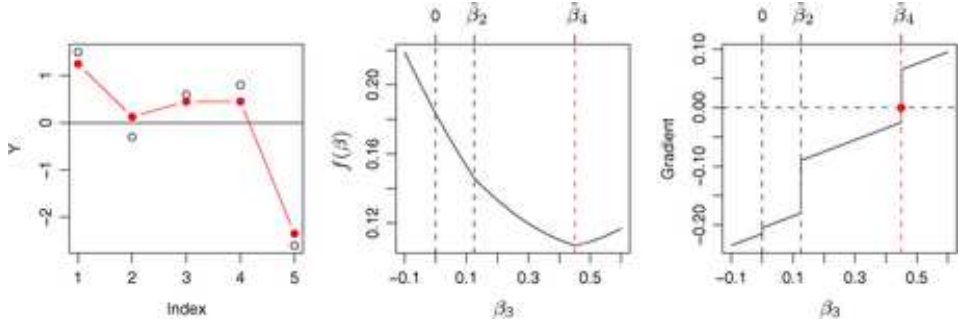

FIG. 6. *Example of the one-dimensional search in the coordinate descent cycle for a FLSA problem with 5 parameters. Shown is the gradient for $\beta_3$, with the other 4 parameters set at the global minimizing values. There are discontinuities at $\tilde{\beta}_2, \tilde{\beta}_4$ and zero. We look for a zero-crossing in each of the intervals $(-\infty, \tilde{\beta}_4), (\tilde{\beta}_4, 0), (0, \tilde{\beta}_2), (\tilde{\beta}_2, \infty)$, and if none is found, take the minimum of $f(\beta)$ over the set of discontinuities. In this case, the minimum is at a discontinuity, with $\tilde{\beta}_3 = \tilde{\beta}_4$.*

*Smoothing cycle*: Here we increase the penalty $\lambda_2$ a small amount, and rerun the two previous cycles.

We now describe each in more detail.

*Descent cycle.*   Consider the derivative of (23), holding all $\beta_k = \tilde{\beta}_k, k \neq i$ fixed at their current values:

$$\frac{\partial f(\beta)}{\partial \beta_i} = -(y_i - \beta_i) + \lambda_1 \cdot \mathrm{sign}(\beta_i)$$
$$- \lambda_2 \cdot \mathrm{sign}(\tilde{\beta}_{i+1} - \beta_i) + \lambda_2 \cdot \mathrm{sign}(\beta_i - \tilde{\beta}_{i-1}),$$

assuming that $\beta_i \notin \{0, \tilde{\beta}_{i-1}, \tilde{\beta}_{i+1}\}$.

The algorithm for coordinate-wise minimization of $f(\beta)$ works as follows. The expression in (24) is piecewise linear in $\beta_i$ with breaks at $0, \tilde{\beta}_{i-1}$ and $\tilde{\beta}_{i+1}$. Suppose, for example, at a given iteration we have $0 \leq \tilde{\beta}_{i-1} \leq \tilde{\beta}_{i+1}$. We check for a zero of (24) in each of the four intervals $(-\infty, 0], (0, \tilde{\beta}_{i-1}], (\tilde{\beta}_{i-1}, \tilde{\beta}_{i+1}]$ and $(\tilde{\beta}_{i+1}, \infty)$. Since the function is piecewise linear, there is an explicit solution, when one exists. If no solution is found, we examine the three active-constraint values for $\beta_i$: $0, \tilde{\beta}_{i-1}$ and $\tilde{\beta}_{i+1}$, and find the one giving the smallest value of the objective function $f(\beta)$. Figure 6 illustrates the procedure on a simulated example.

Other orderings among $0, \tilde{\beta}_{i-1}$ and $\tilde{\beta}_{i+1}$ are handled similarly, and at the endpoints $i = 1$ and $i = p$, there are only three intervals to check, rather than four.

*Fusion cycle.*   The descent cycle moves parameters one at a time. Inspection of Figure 4 shows that this approach can get stuck. One way to get around this is to consider a potential fusion of parameters, when a move of a single $\beta_i$ fails to improve the loss criterion. This amounts to enforcing $|\beta_i - \beta_{i-1}| = 0$ by setting $\beta_i = \beta_{i-1} = \gamma$. With this constraint, we try a descent move in $\gamma$. Equation (24) now becomes

$$
\begin{aligned}
\frac{\partial f(\beta)}{\partial \gamma} = &-(y_{i-1} - \gamma) - (y_i - \gamma) \\
&+ 2\lambda_1 \cdot \text{sign}(\gamma) - \lambda_2 \cdot \text{sign}(\tilde{\beta}_{i+1} - \gamma) \\
&+ \lambda_2 \cdot \text{sign}(\gamma - \tilde{\beta}_{i-2}).
\end{aligned}
\tag{25}
$$

If the optimal value for $\gamma$ decreases the criterion, we accept the move setting $\beta_i = \beta_{i-1} = \gamma$.

Notice that the fusion step is equivalent to temporarily collapsing the problem to one with $p - 1$ parameters:

- we replace the pair $y_{i-1}$ and $y_i$ with the average response $\bar{y} = (y_{i-1} + y_i)/2$ and an observation weight of 2;
- the pair of parameters $\beta_{i-1}$ and $\beta_i$ are replaced by a single $\gamma$, with a penalty weight of 2 for the first penalty.

At the end of the entire process of descent and fusion cycles for a given $\lambda_2$, we identify adjacent nonzero solution values that are equal and collapse the data accordingly, maintaining a weight vector to assign weights to the observations averages and the contributions to the first penalty.

*Smoothing cycle.* Although the fusion step often leads to a decrease in $f(\tilde{\beta})$, it is possible to construct examples where, for a particular value of $\lambda_2$, no fusion of two neighbors causes a decrease, but a fusion of three or more can. Our final strategy is to solve a series of fused lasso problems sequentially, fixing $\lambda_1$, but varying $\lambda_2$ through a range of values increasing in small increments $\delta$ from 0.

The smoothing cycle is then as follows:

1. Start with $\lambda_2 = 0$, hence, with the lasso solution with penalty parameter $\lambda_1$.
2. Increment $\lambda_2 \leftarrow \lambda_2 + \delta$, and run the descent and fusion cycles repeatedly until no further changes occur. After convergence of the process for a given value $\lambda_2$, identify neighboring solution values that are equal and nonzero and collapse the problem as described above, updating the weights.
3. Repeat step 2 until a target value of $\lambda_2$ is reached (or a target bound $s_2$).

Our strategy relies on the following assumptions:

(A1) *If the increments $\delta$ are sufficiently small, fusions will occur between no more than two neighboring points at a time.*

(A2) *Two parameters that are fused in the solution for* $(\lambda_1, \lambda_2)$ *will be fused for all* $(\lambda_1, \lambda'_2 > \lambda_2)$.

By collapsing the data after each solution, we can achieve long fusions by a sequence of pairwise fusions. Note that each of the fused parameters can represent more than one parameter in the original problem. For example, if $\beta_j$ has a weight of 3, and $\beta_{i+1}$ a weight of 2, then the merged parameter has a weight of 5, and represents 5 neighboring parameters in the original problem.

After $m$ fusions, the problem has the form

$$(26) \quad C_m + \min_\beta \tfrac{1}{2} \sum_{i=1}^{n-m} w_i(y_i - \beta_i)^2 + \lambda_1 \sum_{i=1}^{n-m} w_i|\beta_i| + \sum_{i=2}^{n-m} |\beta_i - \beta_{i-1}|.$$

Initially, $m = 0$, $w_i = 1$, and $C_0 = 0$. If the $(m+1)$st fusion is between $\beta_{i-1}$ and $\beta_i$, then the following updates occur:

- $\bar{y} \leftarrow (w_{i-1}y_{i-1} + w_i y_i)/(w_{i-1} + w_i)$.
- $w^+ \leftarrow w_{i-1} + w_i$.
- $C_{m+1} = C_m + \tfrac{1}{2}[w_{i-1} \cdot (y_{i-1} - \bar{y})^2 + w_i \cdot (y_i - \bar{y})^2]$.
- $y_{i-1} \leftarrow \bar{y}$, $w_{i-1} \leftarrow w^+$.
- Discard observation $i$, and reduce all indices greater than $i$ by 1.

Note that we don't actually need to carry out the update for $C_m$, because no parameters are involved.

Figure 7 shows an example with just 9 data points. We have fixed $\lambda_1 = 0.01$ and show the solutions for four values of $\lambda_2$. As $\lambda_2$ increases, the number of fused parameters increases.

Assumption (A1) requires that the data have some randomness (i.e., no pre-existing flat plateaus exist). Assumption (A2) holds in general. We prove that both assumptions hold for the FLSA procedure in the next section.

Numerical experiments show that (A2) does not always hold for the general fused lasso. Hence, the extension of this algorithm to the general fused lasso (detailed in the Appendix) is not guaranteed to yield the exact solution. Note that each descent and fusion cycle can only decrease the convex objective and, hence, must converge. We terminate this pair of cycles when the change in parameter estimates is less than some threshold. The smoothing cycle is done over a discrete grid of $\lambda_2$ values.

**4. Optimality conditions.** In this section we derive the optimality conditions for the FLSA problem, and use them to show that our algorithms' assumptions (A1) and (A2) are satisfied.

We consider the Lagrangian form (23) for the fused lasso. The standard Karush–Kuhn–Tucker conditions for this problem are fairly complicated, since we need to express each parameter in terms of its positive and negative parts in order to make the penalty differentiable. A more convenient formulation is through the sub-gradient approach [see, e.g., Bertsekas (1999), Proposition B.24]. The equations for the subgradient have the form

$$
(27)
\begin{aligned}
&-(y_1 - \beta_1) + \lambda_1 s_1 - \lambda_2 t_2 = 0, \\
&-(y_j - \beta_j) + \lambda_1 s_j + \lambda_2 (t_j - t_{j+1}) = 0, \qquad j = 2, \dots, n,
\end{aligned}
$$

with $s_j = \text{sign}(\beta_j)$ if $\beta_j \neq 0$ and $s_j \in [-1, 1]$ if $\beta_j = 0$. Similarly, $t_j = \text{sign}(\beta_j - \beta_{j-1})$ if $\beta_j \neq \beta_{j-1}$ and $t_j \in [-1, 1]$ if $\beta_j = \beta_{j-1}$. These $n$ equations are necessary and sufficient for the solution. We restate assumptions (A1) and (A2) more precisely, and then prove they hold.

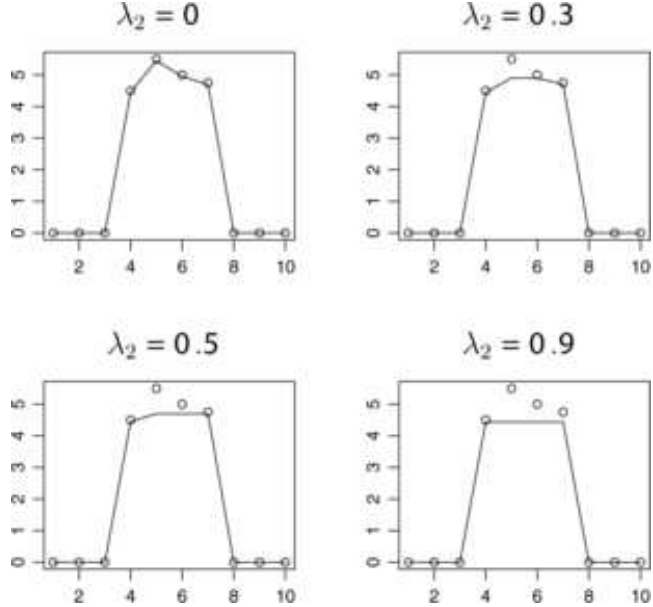

Fig. 7. *Small example of the fused lasso. $\lambda_1$ is fixed at 0.01; as $\lambda_2$ increases, the number of fused parameters increases.*

PROPOSITION 2. *For the fused-lasso signal approximation algorithm detailed in Section 3:*

(A1′) *If the sequence $y_i$ are in general position—specifically, no two consecutive $y_j$ values are equal—and the increments $\delta$ are sufficiently small, fusions will occur between no more than two neighboring points at a time.*

(A2′) *Two parameters that are fused in the solution for $(\lambda_1, \lambda_2)$ will be fused for all $(\lambda_1, \lambda_2' > \lambda_2)$.*

PROOF. We first prove (A2$'$). Suppose we have a stretch of nonzero solutions $\hat{\beta}_{j-k}, \hat{\beta}_{j-k+1}, \ldots, \hat{\beta}_j$ that are equal for some value $\lambda_2$, and $\hat{\beta}_{j-k-1}$ and $\hat{\beta}_{j+1}$ are not equal to this common value. Then $t_{j-k}$ and $t_{j+1}$ each take a value in $\{-1, 1\}$; we denote these boundary values by $T_{j-k}$ and $T_{j+1}$. Although the parameters $t_{j-k+1}, \ldots, t_j$ can vary in $[-1, 1]$ as $\lambda_2$ changes (while the fused group remains intact), the values depend on only the $(j - k + 1)$st through $j$th equations in the system (27), because the boundary values are fixed. Taking pairwise differences, and using the fact that $\hat{\beta}_{j-k} = \hat{\beta}_{j-k+1} = \cdots = \hat{\beta}_j$, this subgroup of equations simplifies to

$$
\begin{pmatrix}
2 & -1 & 0 & 0 & \cdots & 0 \\
-1 & 2 & -1 & 0 & \cdots & 0 \\
\vdots & \vdots & \vdots & \vdots & \ddots & \vdots \\
0 & 0 & \cdots & -1 & 2 & -1 \\
0 & 0 & \cdots & 0 & -1 & 2
\end{pmatrix}
\begin{pmatrix}
t_{j-k+1} \\
t_{j-k+2} \\
\vdots \\
t_{j-1} \\
t_j
\end{pmatrix}
$$

(28)

$$
= \frac{1}{\lambda_2}
\begin{pmatrix}
y_{j-k+1} - y_{j+k} \\
y_{j-k+2} - y_{j-k+1} \\
\vdots \\
y_{j-1} - y_{j-2} \\
y_j - y_{j-1}
\end{pmatrix}
+
\begin{pmatrix}
T_{j-k} \\
0 \\
\vdots \\
0 \\
T_{j+1}
\end{pmatrix}.
$$

Write this system symbolically as $Mt = \frac{1}{\lambda_2}\Delta y + T$, and let $C = M^{-1}$. The explicit form for $C$ given in Schlegel (1970) gives $C_{\ell 1} = (n - \ell + 1)/(n + 1), C_{\ell n} = \ell/(n + 1)$. It is easy to check for all three possibilities for $T$ that $CT \in [-1, 1]$ elementwise. We know that $t = (\frac{1}{\lambda_2}C\Delta y + CT) \in [-1, 1]$ elementwise as well, since $t$ is a solution to (23) at $\lambda_2$. For $\lambda_2' > \lambda_2$, the elements of the first terms shrink, and hence the values $t(\lambda_2')$ remain in $[-1, 1]$. This implies that the fused set remains fused as we increase $\lambda_2$. These equations describe the path $t(\lambda_2)$ for $\lambda_2$ increasing, and only change when one of the boundary points (fused sets) is fused with the current set, and the argument is repeated. This proves (A2$'$).

We now address (A1$'$). Suppose the data are in general position (e.g., have a random noise component), and we have the lasso-solution $\hat{\beta}_j$ for $\lambda_1$. Because of the randomness, no neighboring nonzero parameters $\hat{\beta}_j$ will be exactly the same. This means for each nonzero value $\hat{\beta}_j$, we can write an equation of the form (27) where we know *exactly* the values for $s_j, t_j$ and $t_{j+1}$ (each will be one of the values $\{-1, +1\}$). This means that we can calculate exactly the path of each such $\beta_j$ as we increase $\lambda_2$ from zero, until an event occurs that changes the $s_j, t_j$. By looking at all pairs, we can identify the time of the first fusion of such pairs. The data are then fused together and reduced, and the problem is repeated. Fusions occur one-at-a-time in this fashion, at a distinct sequence of values for $\lambda_2$. Hence, for $\delta$ small enough in our smoothing step, we can ensure that we encounter these fusions one at a time.   □

TABLE 1
*Run times (CPU seconds) for lasso problems of various sizes n, p and different correlation between the features. Methods are the coordinate-wise optimization (Fortran), LARS (R and Fortran versions) and lasso2 (C language)—the homotopy procedure of* Osborne et al. (2000)

| Method | Population correlation between features | | | | | |
|---|---|---|---|---|---|---|
| | $n = 100, p = 1000$ | | | | | |
| | 0 | 0.1 | 0.2 | 0.5 | 0.9 | 0.95 |
| coord-Fort | 0.31 | 0.33 | 0.40 | 0.57 | 1.20 | 1.45 |
| LARS-R | 2.18 | 2.46 | 2.14 | 2.45 | 2.37 | 2.10 |
| LARS-Fort | 2.01 | 2.09 | 2.12 | 1.947 | 2.50 | 2.22 |
| lasso2-C | 2.42 | 2.16 | 2.39 | 2.18 | 2.01 | 2.71 |
| | $n = 100, p = 5000$ | | | | | |
| | 0 | 0.1 | 0.2 | 0.5 | 0.9 | 0.95 |
| coord-Fort | 4.66 | 4.51 | 3.14 | 5.77 | 4.44 | 5.43 |
| LARS-R | 28.40 | 27.34 | 24.40 | 22.32 | 22.16 | 22.75 |
| LARS-Fort would not run | | | | | | |
| lasso2 would not run | | | | | | |
| | $n = 100, p = 20,000$ | | | | | |
| | 0 | 0.1 | 0.2 | 0.5 | 0.9 | 0.95 |
| coord-Fort | 7.03 | 9.34 | 8.83 | 10.62 | 27.46 | 40.37 |
| LARS-R | 116.26 | 122.39 | 121.48 | 104.17 | 100.30 | 107.29 |
| LARS-Fort would not run | | | | | | |
| lasso2 would not run | | | | | | |
| | $n = 1000, p = 100$ | | | | | |
| | 0 | 0.1 | 0.2 | 0.5 | 0.9 | 0.95 |
| coord-Fort | 0.03 | 0.04 | 0.04 | 0.04 | 0.06 | 0.08 |
| LARS-R | 0.42 | 0.41 | 0.40 | 0.40 | 0.40 | 0.40 |
| LARS-Fort | 0.30 | 0.24 | 0.22 | 0.23 | 0.23 | 0.28 |
| lasso2-C | 0.73 | 0.66 | 0.69 | 0.68 | 0.69 | 0.70 |
| | $n = 5000, p = 100$ | | | | | |
| | 0 | 0.1 | 0.2 | 0.5 | 0.9 | 0.95 |
| coord-Fort | 0.16 | 0.15 | 0.14 | 0.16 | 0.15 | 0.16 |
| LARS-R | 1.02 | 1.03 | 1.02 | 1.04 | 1.02 | 1.03 |
| LARS-Fort | 1.07 | 1.09 | 1.10 | 1.10 | 1.10 | 1.08 |
| lasso2-C | 2.91 | 2.90 | 3.00 | 2.95 | 2.95 | 2.92 |

**5. Comparison of run times.**   In this section we compare the run times of the coordinate-wise algorithm to standard algorithms, for both the lasso and diagonal fused lasso (FLSA) problems. All timings were carried out on a Intel Xeon 2.80GH processor.

5.1. *Lasso speed trials.* We generated Gaussian data with $n$ observations and $p$ predictors, with each pair of predictors $X_j, X_{j'}$ having the same population correlation $\rho$. We tried a number of combinations of $n$ and $p$, with $\rho$ varying from zero to 0.95. The outcome values were generated by

$$(29) \qquad Y = \sum_{j=1}^{p} \beta_j X_j + k \cdot Z,$$

where $\beta_j = (-1)^j \exp(-2(j-1)/20)$, $Z \sim N(0,1)$ and $k$ is chosen so that the signal-to-noise ratio is 3.0. The coefficients are constructed to have alternating signs and to be exponentially decreasing.

Table 1 shows the average CPU timings for the coordinatewise algorithm, two versions of the LARS procedure and lasso2, an implementation of the homotopy algorithm of Osborne et al. (2000). All algorithms are implemented as R language functions. The coordinate-wise algorithm does all of its numerical work in Fortran, while lasso2 does its numerical work in C. LARS-R is the "production version" of LARS (written by Efron and Hastie), doing much of its work in R, calling Fortran routines for some matrix operations. LARS-Fort (due to Ji Zhu) is a version of LARS that does all of its numerical work in Fortran. Comparisons between different programs are always tricky: in particular, the LARS procedure computes the entire path of solutions, while the coordinate-wise procedure and lasso2 solve the problems for a set of pre-defined points along the solution path. In the orthogonal case, LARS takes $\min(n, p)$ steps: hence, to make things roughly comparable, we called the latter two algorithms to solve a total of $\min(n, p)$ problems along the path.

Not surprisingly, the coordinate-wise algorithm is fastest when the correlations are small, and gets slower when they are large. It seems to be very competitive with the other two algorithms in general, and offers some potential speedup, especially when $n > p$.

Figure 8 shows the CPU times for coordinate descent, for the same problem as in Table 1. We varied $n$ and $p$, and averaged the times over five runs. We see that the times are roughly linear in $n$ and in $p$.

A key to the success of the coordinate-wise algorithm for lasso is the fact that, for squared error loss, the ingredients needed for each coordinate step can be easily updated as the algorithm proceeds. We can write the second term in (9) as

$$(30) \qquad \sum_{i=1}^{n} x_{ij}(y_i - \tilde{y}_i) = \langle x_j, y \rangle - \sum_{k:\tilde{\beta}_k > 0} \langle x_j, x_k \rangle \tilde{\beta}_k,$$

where $\langle x_j, y \rangle = \sum_{i=1}^{n} x_{ij} y_i$, and so on. Hence, we need to compute inner products of each feature with $y$ initially, and then each time a feature $x_k$ enters the model, we need to compute its inner product with all the rest. But importantly, $O(n)$ calculations do not have to be made at every step. This is the case for all penalized procedures with squared error loss.

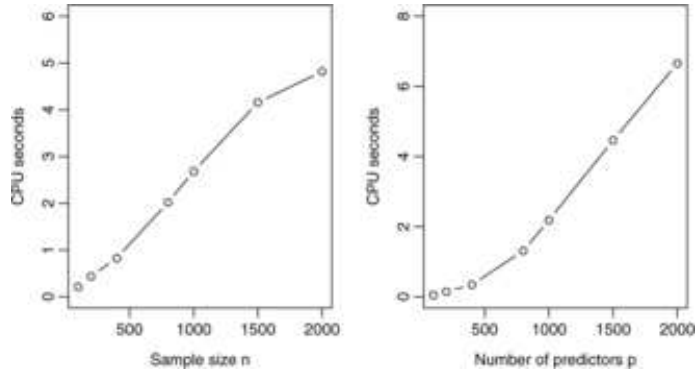

FIG. 8.   *CPU times for coordinate descent, for the same problem as in Table 1, for dif-ferent values of n and p. In each case the times are averaged over five runs and averaged over the set of values of the other parameter (n or p).*

Friedlander and Saunders (2007) do a thorough comparison of the LARS (homotopy) procedure to a number of interior point QP procedures for the lasso problem, and find that LARS is generally much faster. Our finding that coordinate descent is very competitive with LARS therefore suggests that also will outperform interior point methods.

Finally, note that there is another approach to solving the FLSA problem for $\lambda_1 = 0$. We can transform to parameters $\theta_j = \beta_j - \beta_{j-1}$, and we get a new lasso problem in these new parameters. One can use coordinate descent to solve this lasso problem, and then Proposition 1 gives the FLSA solution for other values of $\lambda_1$. However, this new lasso problem has a dense data matrix, and hence, the coordinate descent procedure is many times slower than the procedure described in this section. The procedure developed here exploits the near-diagonal structure of the problem in the original parametrization.

5.2. *Fused lasso speed trials.* For the example of Figure 5, we compared the pathwise coordinate algorithm to the two-phase active set algorithm `sqopt` of Gill, Murray and Saunders (1999). Both algorithms are implemented as R functions, but do all but the setup and wrapup computations in Fortran. Table 2 shows the timings for the two algorithms for a range of values of $\lambda_1$ and $\lambda_2$. The resulting number of active constraints (i.e., $\beta_j = 0$ or $\beta_j - \beta_{j-1} = 0$) is also shown. In the second part of the table, we increased the sample size to 5000. We see that the coordinate algorithm offers substantial speedups, by factors of 50 up to 300 or more.

In these tables, each entry for the pathwise coordinate procedure is the computation time for the entire path of solutions leading to the given val-ues $\lambda_1, \lambda_2$. In practice, one could obtain all of the solutions for a given $\lambda_1$ from a single run of the algorithm, and hence, the numbers in the table are very conservative. But we reported the results in this way to make a fair comparison with the standard procedure since it can also exploit warm starts.

In Table 3 we show the run times for pathwise coordinate optimization for larger values of $n$. As in the previous table, these are the averages of run times for the entire path of solutions for a given $\lambda_1$, and hence, are conservative. We were unable to run the standard algorithm for these cases.

**6. The two-dimensional fused lasso.** Suppose we have a total of $n^2$ cells, laid out in a $n \times n$ grid (the square grid is not essential). We can generalize the diagonal fused lasso (FLSA) to two-dimensions as follows:

$$\min_{\beta} \frac{1}{2} \sum_{i=1}^{n} \sum_{i'=1}^{n} (y_{ii'} - \beta_{ii'})^2$$

TABLE 2
*Run times (CPU seconds) for fused lasso (FLSA) problems of various sizes $n$ for different values of the regularization parameters $\lambda_1, \lambda_2$. The methods compared are the pathwise coordinate optimization, and "standard"-two-phase active set algorithm* `sqopt` *of Gill, Murray and Saunders (1999). The number of active constraints in the solution is shown in each case*

| $\lambda_1$ | $\lambda_2$ | # Active | Coord | Standard |
|---|---|---|---|---|
| | | $n = 1000$ | | |
| 0.00 | 0.01 | 456 | 0.040 | 2.100 |
| 0.00 | 1.00 | 934 | 0.024 | 0.931 |
| 0.00 | 2.00 | 958 | 0.019 | 0.987 |
| 1.00 | 0.01 | 824 | 0.022 | 1.519 |
| 1.00 | 1.00 | 975 | 0.024 | 1.561 |
| 1.00 | 2.00 | 981 | 0.023 | 1.404 |
| 2.00 | 0.01 | 861 | 0.023 | 1.499 |
| 2.00 | 1.00 | 983 | 0.023 | 1.418 |
| 2.00 | 2.00 | 991 | 0.018 | 1.407 |
| | | $n = 5000$ | | |
| 0.000 | 0.002 | 4063 | 0.217 | 20.689 |
| 0.000 | 0.200 | 3787 | 0.170 | 26.195 |
| 0.000 | 0.400 | 4121 | 0.135 | 29.192 |
| 0.200 | 0.002 | 4305 | 0.150 | 41.105 |
| 0.200 | 0.200 | 4449 | 0.141 | 48.998 |
| 0.200 | 0.400 | 4701 | 0.129 | 45.136 |
| 0.400 | 0.002 | 4301 | 0.108 | 41.062 |
| 0.400 | 0.200 | 4540 | 0.123 | 41.755 |
| 0.400 | 0.400 | 4722 | 0.119 | 38.896 |

TABLE 3
*Run times (CPU seconds) for pathwise coordinate optimization applied to fused lasso (FLSA) problems with a large number of parameters $n$ averaged over different values of the regularization parameters $\lambda_1, \lambda_2$*

| $n$ | Average CPU sec |
|---|---|
| 100,000 | 3.54 |
| 500,000 | 14.93 |
| 1,000,000 | 29.81 |

$$
\text{(31)} \quad \text{subject to} \quad \sum_{i=1}^{n} \sum_{i'=1}^{n} |\beta_{ii'}| \leq s_1,
$$

$$
\sum_{i=1}^{n} \sum_{i'=2}^{n} |\beta_{i,i'} - \beta_{i,i'-1}| \leq s_2,
$$

$$
\sum_{i=2}^{n} \sum_{i'=1}^{n} |\beta_{i,i'} - \beta_{i-1,i'}| \leq s_3.
$$

The penalties encourage the parameter map $\beta_{ii'}$ to be both sparse and spatially smooth.

The fused lasso is related to signal processing methods such as "total variation denoising" [Rudin et al. (1992)], which uses a continuous smoothness penalty analogous to the second penalty in the fused lasso. The TV criterion is written in the form

$$
\text{(32)} \qquad \min_{u} \int_{\Omega} |\nabla u|\, du \quad \text{subject to} \quad \|u - y\|^2 = \sigma^2,
$$

where $y$ is the data, $u$ is the approximation with allowable error $\sigma^2$, $\Omega$ is a bounded convex region in $R^d$, $|\cdot|$ denotes Euclidean norm in $R^d$ and $\|\cdot\|$ denotes the norm on $L^2(\Omega)$. Thus, in $d = 1$ dimension, this is a continuous analogue of the fused lasso, but without the (first) $L_1$ penalty on the coefficients. In $d = 2$ dimensions, the TV approach is different: the discretized version uses the Euclidean norm of the first differences in $u$, rather than the sum of the absolute values of the first differences.

This problem (32) can be solved by a general purpose quadratic-programming algorithm; we give details in the Appendix. However, for a $p \times p$ grid, there are $7p^2$ variables and $3p^2 + 3$ constraints, in addition to nonnegativity constraints on the variables. For $p = 256$, for example, this is $458,752$ variables and $196,611$ constraints, so that finding the exact solution is impractical.

Hence, we focus on the pathwise coordinate algorithm. The algorithm has the same form as in the one-dimensional case, except that rather than checking the three active constraint values 0, $\beta_{j-1}$ and $\beta_{j+1}$, we check 0 and the four values to the right, left, above and below (its four-neighborhood). The number of constraint values reduces to 4 at the edges and 3 at the corners. The algorithm starts with individual pixels as the groups, and the four-neighborhood pixels are its "distance-1 neighbors." In each fusion cycle we try to fuse a group with its distance-1 neighbors. If the fusion is accepted, then the distance-1 neighbors of the fused group are the union of the distance-1 neighbors of the two groups being joined (with the groups themselves removed). Now one pixel might be the distance-1 neighbor to each of the two groups being fused, and some careful bookkeeping is required to keep track of this through appropriate weights. Full details are given in the Appendix.

We do not provide a proof of the correctness of this procedure. However, in our (limited) experiments we have found that it gives the exact solution to the two-dimensional fused lasso. We guess that a proof along the lines of that in the one-dimensional case can be constructed, although some additional assumptions may be required.

As in the one-dimensional FLSA (Proposition 1), if we write the problem in terms of Lagrange multipliers $(\lambda_1, \lambda_2, \lambda_3)$, the solution for $(\lambda'_1 > \lambda_1, \lambda_2, \lambda_3)$ can be obtained by soft-thresholding the solutions for $(\lambda_1, \lambda_2, \lambda_3)$.

6.1. *Example* 1. Figure 9 shows a toy example. The data are in the top left, representing a "+"-shaped image with $N(0,1)$ noise added. The reconstruction by the lasso and fused lasso are shown in the other panels. In each case we did a grid search over the tuning parameters using a kind of two-fold validation. We created a training set of the odd pixels ($1, 3, 5 \ldots$ in each direction) and tested it on the even pixels. For illustration only, we chose the values that minimized the squared reconstruction error over the test set. We see that the fused lasso has successfully exploited the spatial smoothness and provided a much better reconstruction than the lasso.

Table 4 shows the number of CPU seconds required for the standard and pathwise coordinate descent algorithms, as $n$ increases. We were unable to apply the standard algorithm for $n = 256$ (due to memory requirements), and have instead estimated the time by crude quadratic extrapolation.

6.2. *Example* 2. Figure 10 shows another example. The noiseless image (top panel) was randomly generated with $512 \times 512$ pixels. The background pixels are zero, while the signal blocks have constant values randomly chosen between 1 and 4. The top right panel shows the reconstruction by the fused lasso: as expected, it is perfect. In the bottom left we have added Gaussian noise with standard deviation 1.5. The reconstruction by the fused lasso is shown in the bottom right panel, using two-fold validation to estimate $\lambda_1, \lambda_2$. The reconstruction is still quite good, capturing most of the important features. In this example we did a search over 10 $\lambda_1$ values. The entire

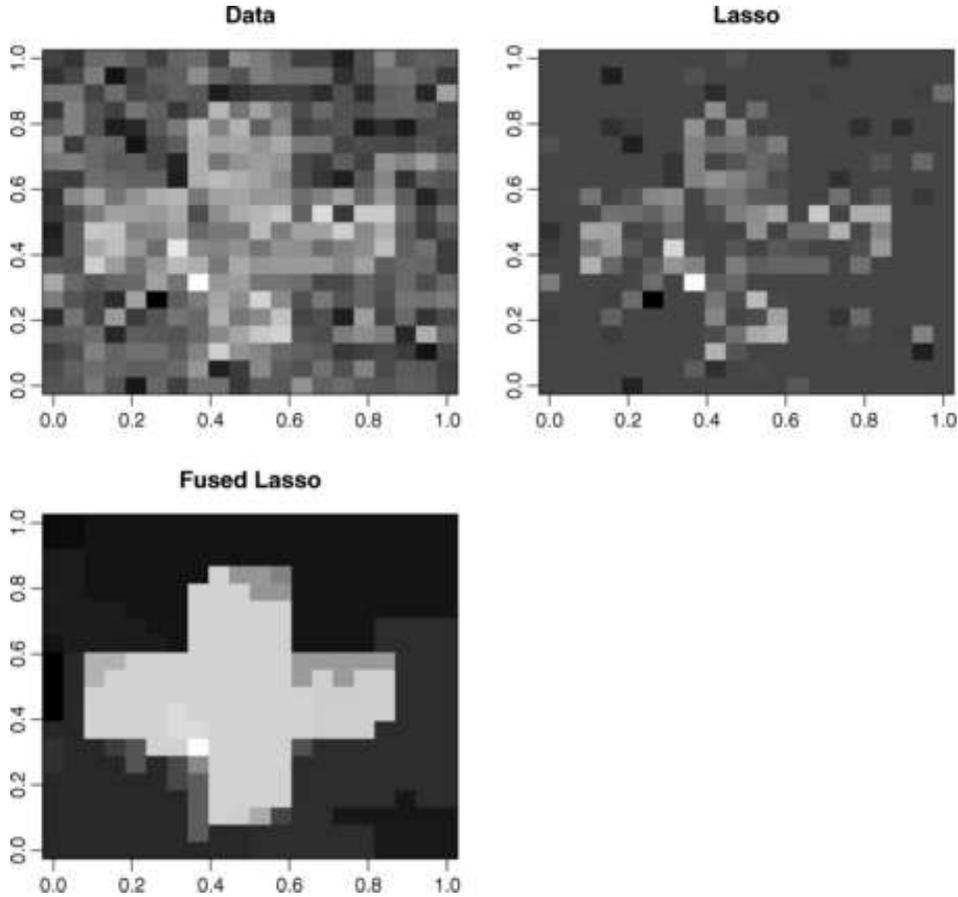

FIG. 9. *A toy example: the data are in the top left, representing a "+"-shaped image with added noise. The reconstructions by the lasso and fused lasso are shown in the other panels. In each case we did a grid search over the tuning parameters using a kind of two-fold validation.*

TABLE 4

*2D fused lasso applied to the toy problem. The table shows the number of CPU seconds required for the standard and pathwise coordinate descent algorithms, as n increases. The regularization parameters were set at the values that yielded the solution in the bottom left panel of Figure 9*

| $n$ | Standard | Coord |
|-----|----------|-------|
| 8 | 2.0 s | 0.07 s |
| 16 | 3.4 s | 0.13 s |
| 32 | 20.8 s | 0.38 s |
| 256 | 38 min | 7.1 s |

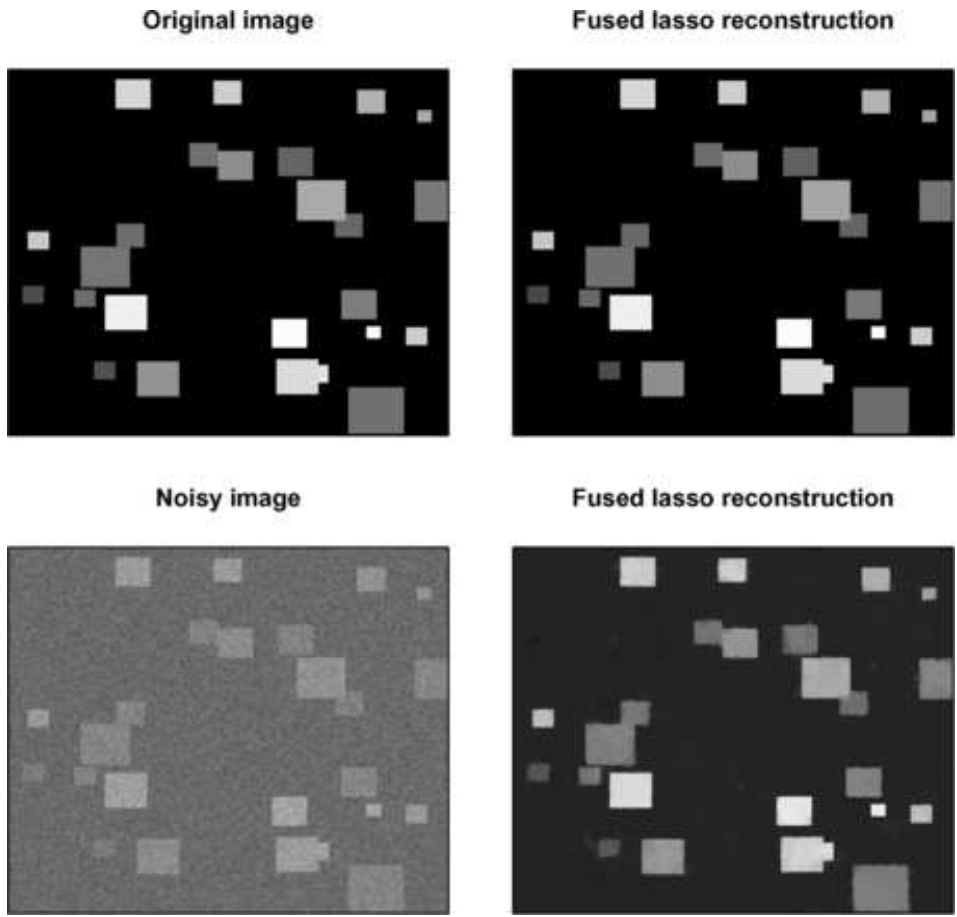

Fig. 10. *A second toy example. The* $100 \times 100$ *noiseless and noisy images are shown on the left, while the corresponding fused lasso reconstructions are shown on the right. In each case we did a grid search over the tuning parameters* $\lambda_1, \lambda_2$ *using a kind of two-fold validation.*

computation for the bottom right panel of Figure 10, including the two-fold validation to estimate the optimal values for $\lambda_1$ and $\lambda_2$, took 11.3 CPU minutes.

6.3. *Example* 3. The top left panel of Figure 11 shows a $256 \times 256$ gray scale image of statistician Sir Ronald Fisher. In the top right we have added Gaussian noise with a standard deviation 2.5. We explore the use of the two-dimensional fused lasso for denoising this image. However, the first (lasso) penalty doesn't make sense here, as zero does not represent a natural baseline. So instead, we tried a pure fusion model, with $\lambda_1 = 0$. We found the best value of $\lambda_2$, in terms of reconstruction error from the original noiseless image. The solution shown in the bottom right panel gives a reasonable approximation to the original image and reduces the reconstruction error from 6.18 to 1.15. In the bottom left panel we have set $\lambda_2 = 0$, and optimized over $\lambda_1$. As expected, this pure lasso solution does poorly, and the optimal value of $\lambda_1$ turned out to be 0.

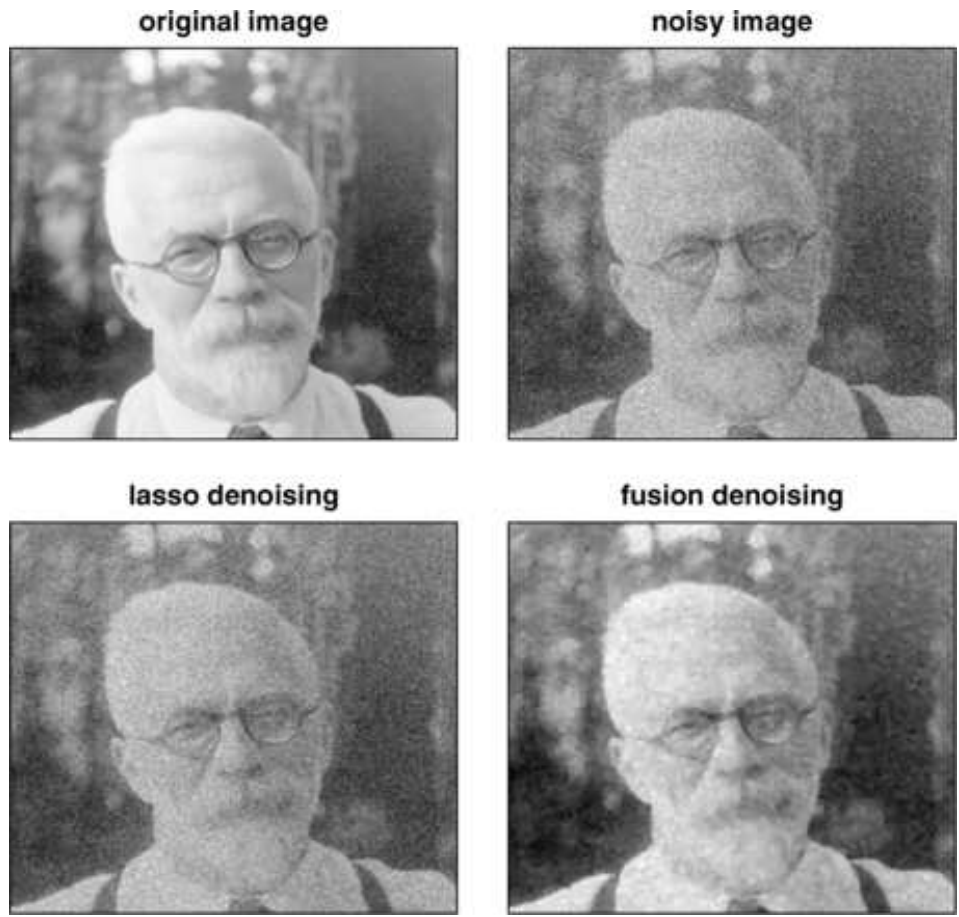

Fig. 11. *Top-left panel: 256 × 256 gray scale image of statistician Sir Ronald Fisher. Top-right panel: Gaussian noise with standard deviation 2.5 has been added. Bottom-left panel: best solution with $\lambda_2$ set to zero (pure lasso penalty); this gives no improvement in reconstruction error. Bottom-right panel: best solution with $\lambda_1$ set to zero (pure fusion penalty). This reduces the reconstruction error from 6.18 to 1.15.*

6.4. *Applications to higher dimensions and other problems.* The general strategy for the two-dimensional fused lasso can be directly applied in higher dimensional problems, the difference being that each cell would have more potential distance-1 neighbors. In fact, the same strategy might be applicable to non-Euclidean problems. All one needs is a notion of distance-1 neighbors and the property that the distance-1 neighbors of a fusion of two groups are the union of the distance-1 neighbors of the two groups, less the fused group members themselves.

**7. Discussion.** Coordinate-wise descent algorithms deserve more attention in convex optimization. They are simple and well-suited to large problems. We have found that for the lasso, coordinate-wise descent is very competitive with the LARS algorithm, probably the fastest procedure for that problem to-date. Coordinate-wise descent algorithms can be applied to problems in which the constraints decouple, and a generalized version of coordinate-wise descent like the one presented here can handle problems in which each parameter is involved in only a limited number of constraints. This procedure is ideally suited for a special case of the fused lasso—the fused lasso signal approximator, and runs many times faster than a standard convex optimizer. On the other hand, it is not guaranteed to work for the general fused lasso problem, as it can get stuck away from the solution.

7.1. *Software.* Both Fortran and R language routines for the lasso, and the one- and two-dimensional fused lasso will be made freely available.

## APPENDIX

**A.1. Proof of Proposition 1.** Suppose that we are optimizing a function of the form

$$f(\beta) = \tfrac{1}{2} \sum_{i=1}^{n} (y_i - \beta_i)^2 + \lambda_1 \sum_{i=1}^{n} |\beta_i| + \sum_{(i,j) \in \mathcal{C}} \lambda_{i,j} |\beta_i - \beta_j|,$$

where $(i,j) \in \mathcal{C}$ if the difference $|\beta_i - \beta_j|$ is $L_1$ penalized with penalty parameter $\lambda_{i,j}$. This general form for the penalty includes the following models discussed earlier:

*Fused lasso signal approximator*: Here, $(i,j) \in \mathcal{C}$ if $i = j - 1$. Furthermore, $\lambda_{i,j} = \lambda_2$.

*Two-dimensional fused lasso*: Here $i$ itself is a two-dimensional coordinate $i = (i_1, i_2)$. Let $(i,j) \in \mathcal{C}$ if $|i_1 - j_1| + |i_2 - j_2| = 1$. If $|i_1 - j_1| = 1$, then $\lambda_{i,j} = \lambda_2$, otherwise $\lambda_{i,j} = \lambda_3$.

Now we prove a soft thresholding result.

LEMMA A.1. *Assume that the solution for $\lambda_1 = 0$ and $\lambda_2 \geq 0$ is known and denote it by $\hat{\beta}(0, \lambda_2)$. Then, the solution for $\lambda_1 > 0$ is*

$$\hat{\beta}_i(\lambda_1, \lambda_2) = \text{sign}(\hat{\beta}_i(0, \lambda_2))(|\hat{\beta}_i(0, \lambda_2)| - \lambda_1)^+.$$

PROOF.   First find the subgradient equations for $\beta_1, \ldots, \beta_n$, which are

$$g_i = -(y_i - \beta_i) + \lambda_1 s_i + \sum_{j\,:\,(i,j)\in\mathcal{C}} \lambda_2 t_{i,j} - \sum_{j\,:\,(j,i)\in\mathcal{C}} \lambda_2 t_{j,i} = 0,$$

where $s_i = \operatorname{sign}(\beta_i)$ if $\beta_i \neq 0$ and $s_i \in [-1, 1]$ if $\beta_i = 0$. Also, $t_{i,j} = \operatorname{sign}(\beta_i - \beta_j)$ for $\beta_i \neq \beta_j$ and $t_{i,j} \in [-1, 1]$ if $\beta_i = \beta_j$. These equations are necessary and sufficient for the solution.

As it is assumed that a solution for $\lambda_1 = 0$ is known, let $s(0)$ and $t(0)$ denote the values of the parameters for this solution. Specifically, $s_i(0) = \operatorname{sign}(\hat{\beta}_i(0))$ for $\hat{\beta}_i(0) \neq 0$ and for $\hat{\beta}_i(0) = 0$, it can be chosen arbitrarily, so set $s_i(0) = 0$. Note that as $\lambda_2$ is constant throughout the whole proof, the dependence of $\beta$, $s$ and $t$ on $\lambda_2$ is suppressed for notational convenience.

In order to find $t(\lambda_1)$, observe that soft thresholding of $\beta(0)$ does not change the ordering of pairs $\hat{\beta}_i(\lambda_1)$ and $\hat{\beta}_j(\lambda_1)$ for those pairs for which at least one of the two is not 0 and, therefore, it is possible to define $t_{i,j}(\lambda_1) = t_{i,j}(0)$. If $\hat{\beta}_i(\lambda_1) = \hat{\beta}_j(\lambda_1) = 0$, then $t_{i,j}$ can be chosen arbitrarily in $[-1, 1]$ and, therefore, let $t_{i,j}(\lambda_1) = t_{i,j}(0)$. Thus, without violating restrictions on $t_{i,j}$, $t(\lambda_1) = t(0)$ for all $\lambda_1 > 0$. $s(\lambda_1)$ will be chosen appropriately below so that the subgradient equations hold.

Now insert $\hat{\beta}_i(\lambda_1) = \operatorname{sign}(\hat{\beta}_i(0))(|\hat{\beta}_i(0)| - \lambda_1)^+$ into the subgradient equations. For $\lambda_1 > 0$, look at 2 cases:

CASE 1.   $|\hat{\beta}_i(0)| \geq \lambda_1$

$$\begin{aligned}
g_i(\lambda_1) &= -y_i + \hat{\beta}_i(0) - \lambda_1 s_i(0) + \lambda_1 s_i(\lambda_1) \\
&\quad + \sum_{j\,:\,(i,j)\in\mathcal{C}} \lambda_2 t_{i,j}(\lambda_1) - \sum_{j\,:\,(j,i)\in\mathcal{C}} \lambda_{j,i} t_{j,i}(\lambda_1) \\
&= -y_i + \hat{\beta}_i(0) + \sum_{j\,:\,(i,j)\in\mathcal{C}} \lambda_2 t_{i,j}(0) - \sum_{j\,:\,(j,i)\in\mathcal{C}} \lambda_{j,i} t_{j,i}(0) = 0
\end{aligned}$$

by setting $s_i(\lambda_1) = s_i(0)$, using the definition of $t(\lambda_1)$ and noting that $\hat{\beta}(0)$ was assumed to be a solution.

CASE 2.   $|\hat{\beta}_i(0)| < \lambda_1$. Here, $\hat{\beta}(\lambda_1) = 0$ and, therefore,

$$\begin{aligned}
g_i(\lambda_1) &= -y_i + \lambda_1 s_i(\lambda_1) + \sum_{j\,:\,(i,j)\in\mathcal{C}} \lambda_2 t_{i,j}(\lambda_1) - \sum_{j\,:\,(j,i)\in\mathcal{C}} \lambda_{j,i} t_{j,i}(\lambda_1) \\
&= -y_i + \hat{\beta}_i(0) + \sum_{j\,:\,(i,j)\in\mathcal{C}} \lambda_2 t_{i,j}(0) - \sum_{j\,:\,(j,i)\in\mathcal{C}} \lambda_{j,i} t_{j,i}(0) = 0
\end{aligned}$$

by choosing $s_i(\lambda_1) = \hat{\beta}_i(0)/\lambda_1 \in [-1, 1]$ and again using that $\hat{\beta}(0)$ is optimal.

As the subgradient equations hold for every $\lambda_1 > 0$, soft thresholding gives the solution. Note that we have assumed that $\lambda_{i,j} = \lambda_2$, but this proof works for any fixed values $\lambda_{i,j}$.

Using this theorem, it is possible to make a more general statement.

PROPOSITION A.1. Let $\hat{\beta}(\lambda_1, \lambda_2)$ be a minimum of $f(\beta)$ for $(\lambda_1, \lambda_2)$. Then the solution for the parameters $(\lambda_1', \lambda_2)$ with $\lambda_1' > \lambda_1$ is a soft thresholding of $\hat{\beta}(\lambda_1, \lambda_2)$, that is,

$$\hat{\beta}(\lambda_1', \lambda_2) = \text{sign}(\hat{\beta}(\lambda_1, \lambda_2))(\hat{\beta}(\lambda_1, \lambda_2) - (\lambda_1' - \lambda_2))^+.$$

PROOF. As a solution for minimizing $f(\beta)$ exists and is unique for all $\lambda_1, \lambda_2 \geq 0$, the solution $\hat{\beta}(0, \lambda_2)$ exists and $\hat{\beta}(\lambda_1, \lambda_2)$ as well as $\hat{\beta}(\lambda_1', \lambda_2)$ are soft-thresholded versions of it using the previous theorem. Therefore,

$$\hat{\beta}_i(\lambda_1, \lambda_2) = \text{sign}(\hat{\beta}_i(0, \lambda_2))(|\hat{\beta}_i(0, \lambda_2)| - \lambda_1)^+,$$
$$\hat{\beta}_i(\lambda_1', \lambda_2) = \text{sign}(\hat{\beta}_i(0, \lambda_2))(|\hat{\beta}_i(0, \lambda_2)| - \lambda_1')^+,$$

for $i = 1, \ldots, n$. If $\hat{\beta}_i(\lambda_1, \lambda_2) = 0$, then also $\hat{\beta}_i(\lambda_1', \lambda_2) = 0$. For $\hat{\beta}_i(\lambda_1, \lambda_2) > 0$, the soft-thresholding implies that the sign did not change and, thus, $\text{sign}(\hat{\beta}_i(\lambda_1, \lambda_2)) = \text{sign}(\hat{\beta}_i(0, \lambda_2))$. It then follows

$$\hat{\beta}_i(\lambda_1', \lambda_2) = \text{sign}(\hat{\beta}_i(0, \lambda_2))(|\hat{\beta}_i(0, \lambda_2)| - \lambda_1')^+$$
$$= \text{sign}(\hat{\beta}_i(\lambda_1, \lambda_2))(|\hat{\beta}_i(\lambda_1, \lambda_2)| - (\lambda_1' - \lambda_1))^+.$$

Therefore, $\hat{\beta}(\lambda_1', \lambda_2)$ is a soft-thresholded version of $\hat{\beta}(\lambda_1, \lambda_2)$. □

*Quadratic programming solution for the two-dimensional fused lasso.* Here we outline the solution to the two-dimensional fused lasso using the general purpose `sqopt` package of Gill, Murray and Saunders (1999).

Let $\beta_{ii'} = \beta_{ii'}^+ - \beta_{ii'}^-$ with $\beta_{ii'}^+, \beta_{ii'}^- \geq 0$. Define $\theta_{ii'}^h = \beta_{i,i'} - \beta_{i-1,i'}$ for $i > 1$, $\theta_{ii'}^v = \beta_{i,i'} - \beta_{i,i'-1}$ for $i' > 1$, and $\theta_{11} = \beta_{11}$. Let $\theta_{ii'}^h = \theta_{ii'}^{h+} - \theta_{ii'}^{h-}$ with $\theta_{i'}^{h+}, \theta_{i'}^{h-} \geq 0$, and similarly for $\theta_{ii'}^v$. We string out each set of parameters into one long vector, starting with the 11 entry in the top left, and going across each row.

Let $L_1$ and $L_2$ be the $n \times n$ matrices that compute horizontal and vertical differences. Let $e$ be a column $n$-vector of ones, and $I$ be the $n \times n$ identity matrix. Then the constraints can be written as

$$
(33) \quad
\begin{pmatrix} -a_0 \\ 0 \\ 0 \\ 0 \end{pmatrix}
\leq
\begin{pmatrix}
L_1 & 0 & 0 & -I & I & 0 & 0 \\
L_2 & 0 & 0 & 0 & 0 & -I & I \\
I & -I & I & 0 & 0 & 0 & 0 \\
0 & e^T & e^T & 0 & 0 & 0 & 0 \\
0 & 0 & 0 & e^T & e^T & 0 & 0 \\
0 & 0 & 0 & 0 & 0 & e_0^T & e_0^T
\end{pmatrix}
\begin{pmatrix} \beta \\ \beta^+ \\ \beta^- \\ \theta^{h+} \\ \theta^{h-} \\ \theta^{v+} \\ \theta^{v-} \end{pmatrix}
\leq
\begin{pmatrix} a_0 \\ 0 \\ 0 \\ s_1 \\ s_2 \\ s_3 \end{pmatrix}.
$$

Here $a_0 = (\infty, 0, 0 \ldots 0)$. Since $\beta_{1i'} = \theta_{1i'}$, setting its bounds at $\pm\infty$ avoids a "double" penalty for $|\beta_{1i'}|$ and similarly for $\beta_{1i'}$. Similarly, $e_0$ equals $e$, with the first component set to zero.

*Pathwise coordinate optimization for the general one-dimensional fused lasso.* This algorithm has exactly the same form as that for the fused lasso signal approximator given earlier. The form of the basic equations is all that changes.

Equation (24) becomes

$$
\begin{aligned}
(34) \quad \frac{\partial f(\beta)}{\partial \beta_j} = -\sum_{i=1}^{n} & \left[ x_{ij} \left( y_i - \sum_{k \neq j}^{p} x_{ik} \tilde{\beta}_k - x_{ij} \beta_j \right) \right] \\
& + \lambda_1 \cdot \text{sign}(\beta_j) - \lambda_2 \cdot \text{sign}(\tilde{\beta}_{j+1} - \beta_j) \\
& + \lambda_2 \cdot \text{sign}(\beta_j - \tilde{\beta}_{j-1}),
\end{aligned}
$$

assuming that $\beta_j \notin \{0, \tilde{\beta}_{j-1}, \tilde{\beta}_{j+1}\}$.

Similarly, expression (25) becomes

$$
\begin{aligned}
(35) \quad \frac{\partial f(\beta)}{\partial \gamma} = -\sum_{i=1}^{n} z_i & \left[ y_i - \sum_{k \notin \{j, j+1\}}^{p} x_{ik} \tilde{\beta}_k - z_i \gamma \right] \\
& + 2\lambda_1 \cdot \text{sign}(\gamma) - \lambda_2 \cdot \text{sign}(\tilde{\beta}_{j+2} - \gamma) \\
& + \lambda_2 \cdot \text{sign}(\gamma - \tilde{\beta}_{j-1}),
\end{aligned}
$$

where $z_i = x_{ij} + x_{i,j+1}$. If the optimal value for $\gamma$ decreases the criterion, we accept the move setting $\beta_j = \beta_{j-1} = \gamma$.

*Pathwise coordinate optimization for two-dimensional fused lasso signal approximator.* Consider a set (grid) $G$ of pixels $p = (i, j)$, with $1 \leq i \leq n_1$, $1 \leq j \leq n_2$. Associated with each pixel is a data value $y_p = y_{ij}$. The goal is to obtain smoothed values $\hat{\beta}_p = \hat{\beta}_{ij}$ for the pixels that jointly solve

$$
\min_{\{\beta_{ij}\}} \frac{1}{2} \sum_{i=1}^{n_1} \sum_{j=1}^{n_2} (y_{ij} - \beta_{ij})^2 + \lambda_1 \sum_{i=1}^{n_1} \sum_{j=1}^{n_2} |\beta_{ij}|
$$

$$(36) \qquad + \lambda_2 \sum_{i=2}^{n_1} \sum_{j=2}^{n_2} |\beta_{ij} - \beta_{i-1,j}| + \sum_{i=1}^{n_1} \sum_{j=1}^{n_2} |\beta_{ij} - \beta_{i,j-1}|.$$

Defining the distance between two pixels $p = (i, j)$ and $p' = (i', j')$ as $d(p, p') = |i - i'| + |j - j'|$, (36) can be expressed as

$$(37) \qquad \min_{\{\beta_p\}} \tfrac{1}{2} \sum_{p \in G} (y_p - \beta_p)^2$$

$$+ \lambda_1 \sum_{p \in G} |\beta_p| + (\lambda_2/2) \sum_{d(p,p')=1} |\beta_p - \beta_{p'}|.$$

Consider a partition of $G$ into $K$ contiguous groups $\{G_k\}$, $G = \bigcup_k G_k$ and $G_k \cap G_{k'} = 0$, $k \neq k'$. A group $G_k$ is contiguous if any $p \in G_k$ can be reached from any other $p' \in G_k$ by a sequence of distance-one steps within $G_k$. Define the distance between two groups $G_k$ and $G_{k'}$ as

$$D(k, k') = \min_{\substack{p \in G_k \\ p' \in G_{k'}}} d(p, p').$$

Suppose one seeks the solution to (37) under the constraints that all pixels in the same group have the same parameter value. That is, for each $G_k$, $\{\beta_p = \gamma_k\}_{p \in G_k}$. The corresponding optimal group parameter values $\hat{\gamma}_k$ are the solution to the unconstrained problem

$$(38) \qquad \min_{\{\gamma_k\}} \tfrac{1}{2} \sum_{k=1}^{K} N_k (\bar{y}_k - \gamma_k)^2$$

$$+ \lambda_1 \sum_{k=1}^{K} N_k |\gamma_k| + (\lambda_2/2) \sum_{D(k,k')=1} w_{kk'} |\gamma_k - \gamma_{k'}|,$$

where $N_k$ is the number of pixels in $G_k$, $\bar{y}_k$ is the mean of the pixel data values in $G_k$, and

$$w_{kk'} = \sum_{p \in G_k} \sum_{p' \in G_{k'}} I[d(p, p') = 1].$$

Note that (38) is equivalent to (37) when all groups contain only one pixel.

Further, suppose that for a given partition one wishes to obtain the solution to (38) with the additional constrain that two adjacent groups $G_l$ and $G_{l'}$, $D(l, l') = 1$, have the same parameter value $\gamma_m$; that is, $\gamma_l = \gamma_{l'} = \gamma_m$, or equivalently, $\{\beta_p = \gamma_m\}_{p \in G_l \cup G_{l'}}$. This can be accomplished by deleting groups $l$ and $l'$ from the sum in (38) and adding the corresponding "fused" group $G_m = G_l \cup G_{l'}$, with $N_m = N_l + N_{l'}$, $\bar{y}_m = (N_l y_l + N_{l'} y_{l'})/N_m$,

$$(39) \qquad \{G_{k'}\}_{D(m,k')=1}$$

$$= \{G_{k'}\}_{D(l,k')=1} \cup \{G_{k'}\}_{D(l',k')=1} - G_l - G_{l'},$$

and $w_{mk'} = w_{lk'} + w_{l'k'}$.

The strategy for solving (36) is based on (38). As with FLSA (Section 3), there are three basic operations: descent, fusion and smoothing. For a fixed value of $\lambda_1$, we start at $\lambda_2 = 0$ and $n_1 \cdot n_2$ groups each consisting of a single pixel. The initial $\lambda_2 = 0$ solution of each $\gamma_k$ is obtained by soft-thresholding $\gamma_k = S(\bar{y}_k, \lambda_1)$ as in (3). Starting at this solution, the value of $\lambda_2$ is incremented by a small amount $\lambda_2 \leftarrow \lambda_2 + \delta$. Beginning with $\gamma_1$, the descent operation solves (38) for each $\gamma_k$ holding all other $\{\gamma_{k'}\}_{k' \neq k}$ at their current values. The derivative of the criterion in (38) with respect to $\gamma_k$ is piecewise linear with breaks at 0, $\{\gamma_{k'}\}_{D(k,k')=1}$. The solution for $\gamma_k$ is thus obtained in the same manner as that for the one–dimensional problem described in Section 3. If this solution fails to change the current value of $\gamma_k$, successive provisional fusions of $G_k$ with each $G_{k'}$ for which $D(k, k') = 1$ are considered, and the solution for the corresponding fused parameter $\gamma_m$ is obtained. The derivative of the criterion in (38) with respect to $\gamma_m$ is piecewise linear with breaks at 0, $\{\gamma_{k'}\}_{D(m,k')=1}$ (39). If any of these fused solutions for $\gamma_m$ improves the criterion, one provisionally sets $\gamma_k = \gamma_{k'} = \gamma_m$. If not, the current value of $\gamma_k$ remains unchanged.

One then applies the descent/fusion strategy to the next parameter, $k \leftarrow k + 1$, and so on until a complete pass over all parameters $\{\gamma_k\}$ has been made. These passes (cycles) are then repeated until one complete pass fails to change any parameter value, at which point the solution for the current value of $\lambda_2$ has been reached.

At this point each current group $G_k$ is permanently fused (merged) with all groups $G_{k'}$ for which $\gamma_k = \gamma_{k'}$, $\gamma_{k'} \neq 0$, and $D(k, k') = 1$, producing a new criterion (38) with potentially fewer groups. The value of $\lambda_2$ is then further incremented $\lambda_2 \leftarrow \lambda_2 + \delta$ and the above process is repeated starting at the solution for the previous $\lambda_2$ value. This continues until a specified maximum value of $\lambda_2$ has been reached or until only one group remains.

**Acknowledgments.** We thank Anita van der Kooij for informing us about her work on the elastic net, Michael Saunders and Guenther Walther for helpful discussions, and Balasubramanian Narasimhan for valuable help with our software. A special thanks to Stephen Boyd for telling us about the subgradient form (28). While writing this paper, we learned of concurrent, independent work on coordinate optimization for the lasso and other convex problems by Ken Lange and Tongtong Wu. We thank the Editors and two referees for comments led to substantial improvements in the manuscript.

J. Friedman
Department of Statistics
Stanford University
Stanford, California 94305
USA
E-mail: jhf@stanford.edu

T. Hastie
Departments of Statistics, and Health
    Research & Policy
Stanford University
Stanford, California 94305
USA
E-mail: hastie@stanford.edu

H. Höfling
Department of Statistics
Stanford University
Stanford, California 94305
USA
E-mail: hhoeflin@gmail.com

R. Tibshirani
Departments of Health,
    Research & Policy and Statistics
Stanford University
Stanford, California 94305
USA
E-mail: tibs@stanford.edu